\documentclass[11pt,fleqn]{article}
\usepackage{amsmath}
\usepackage{amssymb}
\usepackage{mathtools}
\usepackage[T1]{fontenc}
\usepackage[utf8]{inputenc}   % pozwala wprowadzać polskie znaki
\usepackage{color}
\usepackage{url}
\usepackage{hyperref}
\usepackage{aobs-tikz}
\usetikzlibrary{arrows.meta}
\usepackage{graphicx}
\usepackage{float}
\usepackage{authblk}

\DeclareMathOperator{\dip}{dip}

\begin{document}
\date{}
\title{\bf Gauge-invariant quadratic approximation of quasi-local mass and its relation with Hamiltonian
 for gravitational field}
\author[1]{Jacek Jezierski\thanks{Jacek.Jezierski@fuw.edu.pl}}
\author[2]{Jerzy Kijowski\thanks{kijowski@cft.edu.pl}}
\author[2]{Piotr Waluk\thanks{waluk@cft.edu.pl}}
\affil[1]{%Department of Mathematical Methods in Physics,
Faculty of Physics, University of Warsaw, \authorcr
ul. Pasteura 5, 02-093 Warsaw, Poland}
\affil[2]{Center for Theoretical Physics, Polish Academy of Sciences, \authorcr
 Aleja Lotnik\'ow 32/46, 02-668 Warsaw, Poland}

%definicje
\newcommand{\ul}{\underline}

\newcommand{\kolo}[1]{\vphantom{#1}\stackrel{\circ}{#1}\!\vphantom{#1}}
\newcommand{\dtwo}{\kolo{\Delta}}
\newcommand{\ddtwo}{(\dtwo +2 )}
\newcommand{\inv}{^{-1}}
\newcommand{\invmd}{{\mathcal A}}
\newcommand{\Rie}[2]{\mathrm{R}^{#1} {}_{#2}}

\newcommand{\x}{{\bf x}}
\newcommand{\X}{{\bf X}}
\newcommand{\y}{{\bf y}}
\newcommand{\Y}{{\bf Y}}
\newcommand{\B}{{\cal B}}
\newcommand{\g}{{\rm g}}
\newcommand{\T}{{\rm T}}
\newcommand{\N}{{\rm N}}
\renewcommand{\P}{{\rm P}}

\newcommand{\dd}[1][]{\mathrm{d}{#1}}
\newcommand{\arsh}{\mathop {\rm arsinh}\nolimits }
%\newcommand{\be}{\begin{equation}}
%\newcommand{\ee}{\end{equation}}
%\newcommand{\ten}[3]{{#1}^{#2}_{\ #3}}
%\newcommand{\tenud}[3]{{#1}^{#2}{_{#3}}}
%\newcommand{\tendu}[3]{{#1}_{#2}{^{#3}}}
%\newcommand{\tenudu}[4]{{#1}^{#2}{_{#3}}{^{#4}}}
%\newcommand{\tendud}[4]{{#1}_{#2}{^{#3}}{_{#4}}}
%\newcommand{\wek}[2]{{#1}^{#2}}
%\newcommand{\kowek}[2]{{#1}_{#2}}
%\newcommand{\base}[2]{{{\partial}\over {\partial
%{#1}^{#2}}}}
%\newcommand{\diff}[2]{\frac{\partial{#1}}{\partial{#2}}}
%\newcommand{\ssl}[2]{{\, \strut ^{#1}}\! {#2}}
%\newcommand{\ssr}[1]{{\, \strut ^{w}}\! {#1}}
%\newcommand{\ssd}[1]{{\, \strut ^{d}}\! {#1}}
%\newcommand{\ssm}[1]{{\, \strut ^{m}}\! {#1}}
%koniec definicji
%%%%%%%%%%%%%%%%%%%%%%%%%%%%%%%%%%%%%%%%%%%%%%%%%%%%%%%%%%%%%%%%%%%%%%%%%
%%% Margin notes
\newcounter{mnotecount}[section]
\renewcommand{\themnotecount}{\thesection.\arabic{mnotecount}}
\newcommand{\mnote}[1]%{}
{\protect{\stepcounter{mnotecount}}$^{\mbox{\footnotesize  $
      \bullet$\themnotecount}}$ \marginpar{\raggedright\tiny
    $\!\!\!\!\!\!\,\bullet$\themnotecount: #1} }
\newcommand{\JJ}[1]{\mnote{\textbf{JJ:} #1}}

%%%%%%%%%%%%%%%%%%%%%%%%%%%%%%%%%%%%%%%%%%%%%%%%%%%%%%%%%%%%%%%%%%%%%%%%%

\maketitle
\begin{abstract}
Gauge invariant, Hamiltonian formulation of field dynamics within a compact region $\Sigma$ with boundary $\partial \Sigma$ is given for the gravitational field linearized over a Kottler metric. The boundary conditions which make the system autonomous are discussed and the corresponding Hamiltonian functional $\mathcal{H}_\text{\tiny Inv}$ is calculated. It is shown that, under specific boundary conditions, the quasi-local Hawking mass $\mathcal{H}_\text{\tiny Haw}$ reduces to $\mathcal{H}_\text{\tiny Inv}$ in the weak field approximation. This observation is a quasi-local version of the classical Brill--Deser result [D. R. Brill, S. Deser, Ann.Phys.(N.Y.) \textbf{50}, 3 (1968)].
\end{abstract}

Dynamics of the linear theory of gravity can be formulated in terms of two gauge-invariant, non-constrained degrees of freedom. For this purpose one can use e.g. selected components of the (linearized) Weyl tensor \cite{lingrav,praca,potor,WalukCQG2019}. The two degrees of freedom contain the entire information about dynamics: knowing them, the complete field configuration can be uniquely (up to  gauge transformations) reconstructed in a quasi-local way
\footnote{We use the term quasi-local to indicate that recovering the field configuration on a given bounded region $U$ requires knowledge of reduced degrees of freedom on a certain compact superset $K(U)$.}.
The phase space of Cauchy data carries a canonical ADM-symplectic structure. The value of its Hamiltonian functional is uniquely determined by this structure and the field dynamics: it is a positive, quadratic form of gauge invariants \cite{praca,WalukCQG2019}. In case of a flat background, this quantity satisfies an important consistency test: it is equal to the second variation (the leading term in the Taylor expansion) of the total ADM energy, commonly accepted as the Hamiltonian function of the complete, nonlinear theory \cite{BrillDeser}.

Any reasonable definition of a quasi-local mass should also satisfy a quasi-local version of the above test. In the present paper we check the consistency of the Hawking mass with the local energy content of the linear theory for a  general spherically-symmetric background: an arbitrary Kottler metric. Our main result is that the second variation of the Hawking mass assigned to topological two-dimensional (2D) spheres constituting a boundary $\partial \Sigma$ of a compact region $\Sigma$ agrees with the amount of (gauge-independent) field energy of the linearized gravity contained within, {\em modulo} a certain gauge-dependent boundary term. The complete agreement can be obtained if we impose an appropriate gauge condition at the boundary $\partial \Sigma$ which annihilates the undesirable term in the Taylor expansion.

The gauge-dependence of the field energy is not a paradox. We stress that the imposed gauge condition plays a role which is much more fundamental than merely a ``convenient gauge'' used to annihilate unwanted terms in the expansion. Indeed, the 2D surface $\partial \Sigma$ plays a double role in the definition of quasi-local mass $E_{\partial \Sigma}$. The first, obvious one, is to demarcate the region $\Sigma$ whose energy content we want to measure \footnote{In General Relativity Theory, two 3D regions $\Sigma_1$ and $\Sigma _2$ having the same boundary contain the same amount of energy due to the diffeomorphism invariance of the dynamics.}. But the second role, that of a ``reference frame'', is related to the very notion of field energy, which is not a scalar quantity: it is always measured {\em with respect to a reference frame}. In special relativity theory, reference frame can be identified with a vector field, say $T$, which must be a symmetry field of the spacetime geometry. Field evolution consisting in shifting the field configuration along this field becomes an autonomous Hamiltonian system, with the Hamiltonian function provided by the Noether theorem. The same procedure works not only for the {\em total} energy, but also for {\em local} energy contained in a bounded 3D region $\Sigma$, provided appropriate boundary conditions at the boundary of the world tube $\Sigma \times {\mathbb R}$ are satisfied, which assure the adiabatic insulation of its interior from the exterior \footnote{Here, again, two 3D regions $\Sigma_1$ and $\Sigma _2$ having the same boundary contain the same amount of energy due to: 1) ``conservation laws'' satisfied by the Noether energy-momentum tensor and 2) boundary conditions. The latter assure the uniqueness of the evolution.}.

In principle, nothing prevents us from using the same construction for an arbitrary vector field $T$. But the resulting Hamiltonian system is no longer autonomous if $T$ is not a symmetry field. The value of the corresponding Hamiltonian is no longer conserved and cannot be interpreted as the field energy. If $T$ is a combination of time translation, space translation, rotation, boost etc., the resulting Hamiltonian function is a strange combination of energy, momentum, angular momentum, static moment etc., but for a generic $T$ it is difficult to find any reasonable interpretation of such a quantity (see \cite{ptcjjk}).

In General Relativity, Noether theorem does not provide any valuable ``energy density'', but the field evolution can still be interpreted as a Hamiltonian system, provided the interior of the tube $\Sigma \times {\mathbb R}$ is adiabatically insulated by appropriate boundary conditions \cite{kijowskiGRG,DeserJPA}.
Again, it is hard to call ``energy'' a Hamiltonian functional obtained {\em via} such a procedure, unless we choose the field $T$ in a way that eliminates all those unwanted ingredients (like rotation, boost, space translation etc.). To do that, the only ``reference frame'' being at our disposal when defining the quasi-local energy $E_{\partial \Sigma}$ is the surface itself: whenever the extrinsic curvature vector $H^\mu$ is spacelike there is a geometrically preferred timelike vector field $T^\mu$ which is orthogonal to $\partial \Sigma$ and to $H^\mu$, i.e.~satisfying $\g_{\mu\nu} T^\mu H^\nu = 0$.

In the flat Minkowski spacetime this procedure produces a useful, self-parallel vector field $T$ only when the surface $\partial \Sigma$ is sufficiently ``rigid'', i.e. contained in a flat 3D hyperplane. Otherwise, the value of the Hamiltonian function obtained this way cannot be interpreted as a field energy. In particular, one cannot expect any reasonable properties (like positivity, convexity etc.) that usually characterize the energy functional and enable us to prove important properties of the field evolution (e.g.~stability).

It turns out that the above rigidity condition can be generalized to a generic, curved spacetime (see \cite{rigid1,rigid2}, where the existence of the eight-parameter family of rigid spheres has been proved). A natural hypothesis arises that the quasi-local Hamiltonian function, defined by imposing appropriate boundary conditions, can be interpreted as the field energy only for such ``rigid spheres''. The main result of our paper supports this hypothesis. Indeed, the curious gauge condition, which is necessary to obtain equality between the quadratic term in the expansion of the Hawking mass and the field energy of the linear gravity, is just the linearized version of the rigidity condition of the surface $\partial \Sigma$.

\section{Technical setup}

We will work within the framework of the Cauchy problem for the Einstein equation, in the ADM formulation thereof.
The background for linearizaton of the theory will be a Kottler metric:
\begin{equation}\label{Kottler}
\eta_{\mu\nu}\dd x^\mu \dd x^\nu =-f\dd[t]^2 + \frac{1}{f} \dd[r]^2+ r^2\left[\dd[\vartheta]^2+\sin^2\vartheta\dd[\varphi]^2 \right] \, , %\quad
\end{equation}
which is spherically symmetric and $f(r):=1-\frac{2m}{r}-\frac{r^2}{3}\Lambda$.
Our choice of coordinates: $(x^0,x^1,x^2,x^3) = (\,t\,,\,\vartheta\,,\,\varphi\,,\,r\,)$ is fixed by this form of the background metric.
%We order them in the following way: $(x^0,x^1,x^2,x^3) = (\,t\,,\,\vartheta\,,\,\varphi\,,\,r\,)$.

We consider a compact region on a Cauchy surface, foliated by a family of two-dimensional spheres:
\begin{equation}
\Sigma =\{ x^0 = t_0, r_0 \leq x^3 \leq r_1 \} = \underset{r \in [r_0,r_1] }{\bigcup} S(r),  \quad
S(r)= \{ x \in \Sigma : x^3=r  \}.
\end{equation}
We assume that $\Sigma$ lies within the domain of positive $f$ --- outside of the black hole horizon and within the cosmological horizon (if it exists \footnote{Existence and location of horizons is determined by positive roots of the cubic polynomial $r\, f(r)$. For $\Lambda<0$ only one positive root, the black hole radius $r_S$, exists and we assume $r_S \leq r_0 < r_1$. For $\Lambda>0$ there are two positive roots, interpreted as the black hole and cosmological horizon radii $r_S<r_C$. In this case, we take $r_S\leq r_0<r_1\leq r_C$.}).

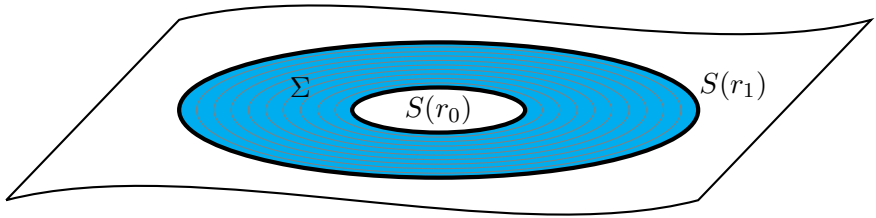
\begin{figure}[H]
\centering
\begin{tikzpicture} [yscale=0.6,xscale=2.3] %[yscale=0.5,xscale=1.7]
\draw[thick] (0,0) -- (1,4) to [out=45,in=225] (5,4) -- (4,0) to [out=225,in=45] (0,0);
\draw[ultra thick, fill=cyan] (2.5,2) circle [radius=1.5];
\draw[ultra thick, fill=white] (2.5,2) circle [radius=0.5];
\foreach \q in {0.6,0.7,...,1.4} \draw[thin,gray] (2.5,2) circle [radius=\q];
\node at (2.5,2) {$S(r_0)$};
\node [above right] at (3.95,2) {$S(r_1)$};
\node at (1.7,2.5) {$\Sigma$};
\end{tikzpicture}
\end{figure}
A following indexing convention will be used to denote dimensionality of geometric objects:
small Greek, small Latin and big Latin indices will denote full spacetime $(0,1,2,3)$, Cauchy surface $(1,2,3)$ and
2D sphere $(1,2)$ coordinates repectively.
%small Greek indices correspond to full spacetime coordinates $(\mu , \nu=0,1,2,3)$, small Latin indices denote objects from the Cauchy surface $(k,l=1,2,3)$ and big Latin indices represent coordinates on two-dimensional spheres $S(r)$ $(A,B=1,2)$.
The coordinate derivative and the two-dimensional covariant derivative on $S(r)$ will be denoted by a comma and the symbol ``$||$''.
The volume form on the spheres defined by the Kottler metric is $\Pi:=r^2 \sin \theta$ and $\lambda$ denotes the counterpart volume form for $\g_{AB}$. The symbol $\dtwo$ is the Laplace--Beltrami operator on a unit sphere.% This notation agrees with the one used in \cite{WalukCQG2019}.
We use a geometric set of units, in which both the speed of light and the gravitational constant are equal to one, $c=1=G$.

%\section{Reduced variables. Hamiltonian version of linearized theory}

The complete description of the canonical structure of linear gravity on the Kottler background has been given in paper \cite{WalukCQG2019} (a generalization of earlier results \cite{praca} to the case with cosmological constant). We summarize crucial results below. Initial (Cauchy) data for Einstein equation can be represented in the form of two symmetric tensors --- the induced metric and the ADM momentum (trace-corrected extrinsic curvature of $\Sigma$):
\begin{equation}
(  \g_{kl}, \mathrm{P}^{kl} ), \quad
\g_{kl}=\g_{\mu\nu}|_{\Sigma}, \quad
\mathrm{P}^{kl}=\sqrt{  \g}( \g^{kl} K - K^{kl} ), \quad
\g:=\det  \g_{kl}.
\end{equation}
In linearized theory, we consider solutions of the form $\g_{\mu\nu} = \eta_{\mu\nu} + h_{\mu\nu}$ and consider corresponding Cauchy data as perturbations of the point $(\eta_{kl},{\rm P}^{kl}(\eta_{\mu\nu})=0)$.
%The perturbation of the ADM momentum will therefore be numerically equal to the ADM momentum of the full metric, and the resulting perturbation data will have the form:
%\begin{equation}
%$(h_{kl},P^{kl})$, $h_{kl}:= {\rm g}_{kl} - \eta_{kl}$, $P^{kl}={\rm P}^{kl}$.
%\end{equation}

The linearized data set $(h_{kl},P^{kl})$ is constrained by four Gauss--Codazzi constraint equations and partially redundant due to a four-parameter family of gauge transformations.
%Furthermore, they are subject to a four-parameter family of gauge transformations --- the ``infinitesimal coordinate changes''. Therefore, out of the twelve independent tensor components, only four carry physically relevant information. We single out these ``true'' degrees of freedom by constructing a set of reduced, gauge invariant variables. To obtain them, we decompose the symmetric tensors with respect to the geometric structure on the two-spheres $S(r)$ and rearrange the parts into suitable combinations. The theory splits into two separate sectors, differing in parity under manifold orientation changes. We call the odd sector ``axial'' and the even one ``polar''. In linear theory these two parts do not interact.
It can be condensed to a set of two pairs of mutually conjugate (like positions and momenta) observables:
%The resulting reduced variables have the following form (see \cite{praca} and \cite{WalukCQG2019}): there are two (mutually conjugate, like position and momentum) observables in the axial sector:
\begin{equation}
\begin{aligned}
 \y&:=2\Pi^{-1}r^2 P^{3A||B}\varepsilon_{AB} \; , %\label{y}
 \\
 \Y&:= \Pi(\dtwo +2)h_{3A||B}\varepsilon^{AB}
 -\Pi (r^2 h^C{_{A||CB}}\varepsilon^{AB})_{,3} %\label{Y}
 \; , \\
 \x& := r^2h_{AB}{^{||AB}}-(\dtwo+1)(h_{AB}\eta^{AB})+ f \B Q
   \; ,               % \label{x}
   \\
 \X &:= 2r^2 P^{AB}{_{||AB}} - \dtwo (P^{AB}\eta_{AB}) + \B  \Xi
\; , %\label{X}
\end{aligned}
\end{equation}
where $\Xi$ and $Q$ denote the following expressions:
\begin{align}
\Xi &:= 2rP^{3A}{}_{||A} +\dtwo P^3{}_3 \, ,\label{Xi_def}\\
Q &:= 2 h^3{}_3 +2r h_3{}^A{}_{||A}-r(h_{AB}\eta^{AB})_{,3} \, , \label{Q_def}
\end{align}
 and we have introduced a following quasi-local (non-local on individual spheres $S(r)$) operator:
\vspace*{-0.2cm}
\begin{equation}
 \B := (\dtwo +2)\left(\dtwo +2-\frac{6m}r\right)^{-1}.
 \vspace*{-0.2cm}
\end{equation}

%The operator $\B$ is non-local with respect to the coordinates on spheres $S(r)$. It is, however, local with respect to the radial and temporal coordinates. The value of $\B$ in a compact subset of spacetime will therefore always depend upon a compact, though perhaps slightly bigger, subset of spacetime. We call such an operator ``quasi-local''.

The set of four scalar functions $(\x,\X,\y,\Y)$:
\begin{enumerate}
\item is gauge invariant,
\item is no longer restricted by any constraint,
\item diagonalizes the ADM symplectic structure of the phase space of Cauchy data (i.e. is a set of {\em canonical} variables),
\item carries the {\em entire} physical information about the gravitational field, i.e.~$(h_{kl},P^{kl})$ can be uniquely reconstructed {\em up to gauge transformations} from this set,
\item reduces field dynamics to a set of four equations of motion:
\end{enumerate}\vspace{-2ex}
\begin{equation}
\begin{aligned}
%\label{px}
\dot\x&=\frac{f}\Pi \X \,  ,  \\
%\label{pX}
\dot\X &=\frac\Pi{r^2} \left\{\left(f{r^2}\x_{,3}\right)_{,3} + \left[\dtwo +f(1-2 \B ) +1 - r^2\Lambda\right] \B \x\right\},  \\ %\label{py}
\dot\y&=\frac{f}\Pi \Y \, , \\
%\label{pY}
 \dot\Y &= \Pi \left\{\left[\frac f {r^2}(r^2\y)_{,3}\right]_{,3} +\frac1{r^2}(\dtwo +2) \y\right\} \, .
\end{aligned}
\label{dynamics}
\end{equation}
\ \\
One subtle point needs to be discussed here: when decomposed into spherical harmonics on $S(r)$, the monopole and dipole parts of $(\x,\X,\y,\Y)$ are not dynamical --- they encode conserved charges. There are up to ten charges in the perturbation, depending on the symmetry of the background metric.
%total mass, center of mass (three charges), linear momentum (three charges) and angular momentum (three charges --- encoded in the dipole part of $\y$, ``$\dip \y$'').
The monopole part of $\x$ is the mass (energy) of the perturbation. For $m=0$ (pure de Sitter background) the dipole parts of $\X$ and $\x$ describe the linear momentum and the static moment (i.e.~information about center of mass) respectively, producing six charges altogether.
When $m \neq 0$ we lose the translational symmetry of the background and these charges vanish.
% the corresponding dipole parts of $Q$ and $\Xi$ replace/substitute those ``linear charges'', but they are no longer gauge-invariant.
Finally, the dipole part of $\y$ (denoted $\dip\y$) describes the angular momentum, providing the last three charges. The remaining mono-dipole parts of the canonical variables vanish identically.

In the regime of weak fields, the first seven charges can be easily eliminated, as the splitting of $\g_{\mu\nu}$ into background and perturbation is not unique. By changing the background to a Kottler metric with a different mass parameter and/or acted upon with a \emph{small} boost or translation we can modify the value of the charges and shift them from the dynamical field $h$ to the background $\eta$. To eliminate the angular momentum, however, one would need to use a background with a~non-vanishing angular momentum (Kerr-de Sitter). For the sake of simplicity we keep the spherical symmetry of the background and describe angular momentum on the level of perturbation.

% Indeed, splitting the metric tensor $g_{\mu\nu}$ into the background $\eta_{\mu\nu}$ and a ``small perturbation'' $h_{\mu\nu}$ is not unique. If the mass, linear momentum and the static moment/center of mass of the resulting field $h_{\mu\nu}$ do not vanish, we can ``change our mind'' and choose as the background a slightly different Kottler metric $\eta$ obtained from the previous one by a small change of the parameter $m$ and by a Lorentz boost and a spacelike translation (both {\em small}), in such a way, that the resulting charges of the corresponding field $h$ vanish. Physically, this means that the seven charges can be shifted from the dynamical field $h$ to the \mbox{background $\eta$}. To eliminate the angular momentum it would be necessary to use a background with a~non-vanishing angular momentum (Kerr-de Sitter). For the sake of simplicity we keep the spherical symmetry of the background and describe angular momentum on the level of perturbation.

The space of Cauchy data of the complete (non-linear) theory is endowed with the canonical symplectic form $\Omega$, known as the ADM structure.
In case of a bounded region $\Sigma$ with boundary $\partial \Sigma$, the symplectic form $\Omega$ contains not only the ADM bulk term $\int_{\Sigma} \delta {\rm P}^{kl} \wedge \delta g_{kl}$, but also an extra boundary term, which makes it gauge invariant (see \cite{ptcjjk,kijowskiGRG}). It turns out
%(see  \cite{WalukCQG2019})
 that for linearized theory only the dynamical, ``mono-dipole-free'' part, denoted by underscored symbols $(\underline\x,\underline\X,\underline\y,\underline\Y)$,
% (also called the ``wave part'')
of the initial data remains in the bulk integral.
%It represents the ``true dynamical variables'' of the theory.
The ADM symplectic structure assumes a canonical form:
%\begin{widetext}
\begin{equation}
\begin{aligned}
%\underline
{ \Omega} &= \frac 1{16 \pi}\int_\Sigma \delta {P}^{kl}\wedge \delta {h}_{kl} \\
 &= \frac 1{16 \pi}
\int_\Sigma \delta \underline{\X}  \wedge \invmd \delta\underline{\x} + \delta\underline{\Y}\wedge \invmd\delta \underline{\y} + \begin{array}{c} \text{\small boundary} \\ \text{\small terms}\end{array} \, ,
\end{aligned}
\end{equation}
%\end{widetext}
where $\invmd:=\dtwo\inv\ddtwo\inv$ and the boundary terms can be killed by appropriate boundary conditions.

Field dynamics \eqref{dynamics}, symbolically represented as $\frac{\partial}{\partial t}$, uniquely defines the gauge-invariant Hamiltonian functional through formula:
%\begin{equation}
$\displaystyle {\Omega}(\frac{\partial}{\partial t},\cdot) = -  \delta \mathcal{H}_\text{\tiny Inv}$,
%\end{equation}
where
%\begin{widetext}
{%\small
\begin{equation}
\begin{aligned}
16\pi &\mathcal{H}{}_\text{\tiny Inv}=\frac 12 \int_{\Sigma} \frac{f}{\Pi}\left[ \underline{\X} \invmd  \underline{\X} + \underline{\Y} \invmd  \underline{\Y} \right]  \\
&
+\frac 12\int_{\Sigma} \frac{\Pi}{r^2} \left[ f(r\underline{\x})_{,3} \invmd  (r\underline{\x})_{,3} + \underline{\x} \frac{r^2}{f} V^{(+)} \invmd  \underline{\x} \right]   \\
& +\frac 12\int_{\Sigma} \frac{\Pi}{r^2} \left[ f(r\underline{\y})_{,3} \invmd  (r\underline{\y})_{,3} + \underline{\y} \frac{r^2}{f} V^{(-)} \invmd  \underline{\y} \right].
\end{aligned}
\end{equation} }
%\end{widetext}
$V^{(+)}$ and $V^{(-)}$ are quasi-local, positive-definite potential operators, whose exact form is irrelevant here  (cf.~\cite{WalukCQG2019}).
%We again refer the reader to \cite{WalukCQG2019} for further details.

To complete the technical introduction, we provide definitions of extrinsic curvature, extrinsic torsion, and Hawking mass:

For a metric manifold $\mathcal{M}$ with a submanifold $\mathcal{S}$, the \emph{extrinsic curvature} ${\rm H}^\mu{}_{AB}$ of $\mathcal{S}$ in $\mathcal{M}$ is defined by:
\begin{equation}
{\rm H}^\mu{}_{AB} X^A Y^B:=\mathcal{P}^\mu\left( \nabla_X Y \right),
\end{equation}
where $X, Y\in T\mathcal{S}$, $\nabla$ is the metric connection on $\mathcal{M}$ and $\mathcal{P}^\mu$ denotes the orthogonal projection onto the space of vectors orthogonal to $\mathcal{S}$, the normal bundle $N\mathcal{S}$.
%Hence, ${\rm H}^\mu(X,Y)$ is a vector in $N\mathcal{S}$.
The trace of ${\rm H}^\mu{}_{AB} $ with respect to the intrinsic metric of $\mathcal{S}$ is called the \emph{mean} extrinsic curvature $\rm H^\mu$ and is a vector in $N\mathcal{S}$.

If $\dim N\mathcal{S}=1$, we identify elements of $N\mathcal{S}$ with scalar functions on $\mathcal{S}$ by choosing a unit normal vector field as a basis in $N\mathcal{S}$.

For $\mathcal{S}$ of codimension two and a spacelike $\rm H^\mu$ we can find $T^\mu \in N \mathcal{S}$
 complementing $\rm H^\mu/||\rm H||$ to an orthonormal basis and define the \emph{extrinsic torsion} $t_A \in  T^*\mathcal{S}$:
\begin{equation}
t_A X^A:=T_\mu\nabla_X\left(\frac{{\rm H}^\mu}{||{\rm H}||}\right),
\end{equation}
%If, however, $\mathcal{S}$ is of codimension two and $\rm H^\mu$ is nonvanishing, then we may choose a unit vector field $T^\mu\in N\mathcal{S}$, so that the pair constitutes an orthogonal basis of the normal bundle. We can then define the quantity known as \emph{extrinsic torsion}:
%\begin{equation}
%t_A X^A:=T_\mu\nabla_X\left(\frac{{\rm H}^\mu}{||{\rm H}||}\right),
%\end{equation}
The symbol $||\rm H||$ denotes the length of the vector $\rm H^\mu$.

The Hawking mass \cite{Hawking, Laszlo} of a region surrounded by a topological 2D sphere $\mathcal{S}$ is defined by:
\begin{equation}
\mathcal{H}_\text{\tiny Haw} := \sqrt{\frac{\text{\small Area} \; \mathcal{S}}{16 \pi}}\left( 1 -\frac{1}{16\pi}\int_{\mathcal{S}} ({\rm H}_\mu {\rm H}^{\mu} + \frac 43 \Lambda) \dd a   \right)\, ,
\label{Hawking}
\end{equation}
where ${\rm H}^\mu$ is the mean extrinsic curvature of the boundary $\mathcal{S}$ in the enveloping four-dimensional spacetime. Formula (\ref{Hawking}) includes a correction term $\frac 43 \Lambda$, which
accounts for the cosmological constant. % We stress its presence, as it is often omitted in literature.

\section{Main result}

Relation between $\mathcal{H}_\text{\tiny Haw}$ and  $\mathcal{H}_\text{\tiny Inv}$ results from the scalar Gauss--Codazzi constraint:
\begin{equation}
 \g \overset{3}{R} -  2\Lambda  \g = \P^{kl}\P_{kl}
 - \frac{1}{2} \P^2 \,  .
\end{equation}
Using the Gauss--Codazzi geometric identities to express the Ricci scalar of $\Sigma$ in terms of objects on the 2D spheres $S(r)$ and integrating over $\Sigma$, we arrive at a following equation\footnote{The assumption for the term $\frac{\sqrt{\g^{33}}}k +\frac{r}2$ to be vanishing is equivalent to ``inverse mean curvature gauge'' which gives positivity of the right-hand side.}:
\begin{equation}
\begin{aligned}
	\int_{\partial \Sigma}  r \lambda\left( \overset{2}{R} -\frac12  k^2  - \frac 23 \Lambda\right)  =
	- \int_\Sigma \underbrace{\left( \frac{\sqrt{\g^{33}}}k +\frac{r}2 \right)}_{\text{\tiny IMC gauge }}
  \left[  (k^2+ \frac 43 \Lambda) w^a \right]_{,a} + \\
\int_\Sigma \frac{\g^{33}}{\lambda} {\left( P_{kl}P^{kl} -\frac 12 P^2 \right)}
+  \int_\Sigma  \lambda \left( k_{AB}k^{AB} -\frac 12 k^2 + \frac 12 \tilde{\g}^{AB} ( \log \g^{33} ), {_A} (\log \g^{33} ), {_B}\right)\, ,
\end{aligned}
\label{ScalarConstraint}
\end{equation}
\noindent
where $k_{AB}$ is the extrinsic curvature of spheres $S(r)$ in $\Sigma$, $\tilde{\g}^{AB}$ is the inverse of $\g_{AB}$, $k:=k_{AB}\tilde{\g}^{AB}$ and $w^a:=\lambda \frac{\g^{3a}}{\g^{33}}$.
%The boundary integral on the left-hand side is reminiscent of the Hawking mass with the cosmological constant correction term, although it is \emph{not} equal to (\ref{Hawking}) yet.
We will transform the left-hand side of \eqref{ScalarConstraint} into the Hawking mass integral \eqref{Hawking} in the following steps.
%To transform the left-hand side of (\ref{ScalarConstraint}) into the Hawking mass integral \eqref{Hawking} several steps must still be taken.
 We begin by performing a second-order approximation of the right-hand side of the equation. Among other resultant terms, a square of the extrinsic curvature of $\Sigma$ will emerge, which we may transfer to the left-hand side.
We obtain an approximate equality:
{\small
\begin{equation}
\begin{aligned}
&r \int_{\partial \Sigma}  \lambda    \Bigg(  \overset{2}{R} -\frac12  \underbrace{(k^2 - (K_{AB}\eta^{AB})^2)}_{\approx {\rm H}_\mu {\rm H}^{\mu}}  - \frac 23 \Lambda \Bigg)
 \, \approx 16\pi \mathcal{H}_\text{\tiny Inv}  %\\ &
 +\int_{\Sigma} \frac{\Pi}{2r^2} \underbrace{\dip(\y) (-\dtwo)^{-1} \dip (\y)}_{\text{square of angular momentum}} +
 \int_{\partial \Sigma} \frac{2m}{r^2} (\Pi -\lambda) \\
 & +\frac 12 \int_{\partial \Sigma} \frac{f\Pi}{r} \left( \underline\y \invmd \underline\y  + \underline\x (\B-1) \invmd \underline\x \right)
  + %\frac 12 \int_{\partial \Sigma}
  \frac{rf}{\Pi} \underline\Xi \invmd \B \underline\Xi %\\ &
 - \frac 12 \int_{\partial \Sigma} \frac{f\Pi}{r} \B \underline{Q} \dtwo\invmd \left[  \underline\x + \frac 14 \dtwo \underline{Q}
 -\frac 12 f(\B-1)\underline{Q} \right].
\end{aligned}
\label{finalResult}
\end{equation} }
This equation is in fact the main result of our paper, but to fully appreciate its meaning, several finishing touches are still necessary.
First, we need to get rid of the problematic expression $\frac{2m}{r^2}\int_{\partial \Sigma}(\Pi - \lambda)$.
%To do this, we observe that we still have some freedom in choosing the radial coordinate on the manifold with metric $\g_{\mu\nu}$.
We do this by observing, that our assumption of existence of a spherical foliation of $\Sigma$ does not specify the way in which we identify its leaves with the leaves of the natural spherical foliation of $\eta_{\mu\nu}$ ---
the way that we compare $\g_{\mu\nu}$ with the background reference metric is not entirely fixed.
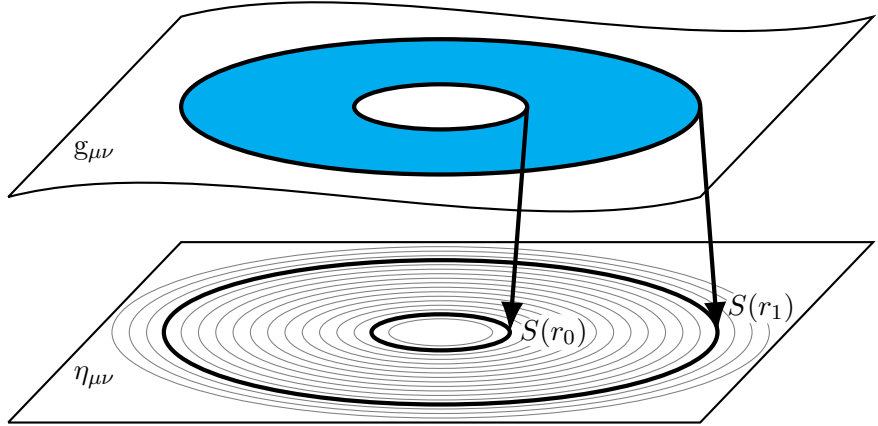
\begin{figure}[H]
\centering
\begin{tikzpicture} [yscale=0.6,xscale=2.3] %[yscale=0.4,xscale=1.6]
\draw[thick] (0,0) -- (1,4) to [out=45,in=225] (5,4) -- (4,0) to [out=225,in=45] (0,0);
\draw[ultra thick, fill=cyan] (2.5,2) circle [radius=1.5];
\draw[ultra thick, fill=white] (2.5,2) circle [radius=0.5];

\draw[thick] (0,-5) -- (1,-1) -- (5,-1) -- (4,-5) -- (0,-5);
\foreach \q in {0.3,0.4,...,1.9} \draw[thin,gray] (2.5,-3) circle [radius=\q];
\draw[ultra thick] (2.5,-3) circle [radius=1.6];
\draw[ultra thick] (2.5,-3) circle [radius=0.4];

\draw [ultra thick, -{Latex[length=4mm]}] (3,2) -- (2.9,-3);
\draw [ultra thick, -{Latex[length=4mm]}] (4,2) -- (4.1,-3);

\draw [color=white ,fill=white] (3.15,-3) circle [x radius=0.2, y radius=0.4];
\draw [color=white ,fill=white] (4.35,-2.4) circle [x radius=0.2, y radius=0.4];
\node [right] at (2.9,-3) {$S(r_0)$};
\node [above right] at (4.1,-3) {$S(r_1)$};
\node at (0.5,1) {$\g_{\mu\nu}$};
\node at (0.5,-4) {$\eta_{\mu\nu}$};
\end{tikzpicture}
\end{figure}
In practice, this gives us freedom of remapping $r$ with a monotonous function.
%Using it, we can make the radial coordinate on the boundary equal to the areal radius for the metric $\g_{\mu\nu}$. As it is already the areal radius for the metric $\eta_{\mu\nu}$, this condition leads to:\\
Using it, we can make the radial coordinate on the boundary equal to the areal radius for both the metric $\eta_{\mu\nu}$ (by definition) and $\g_{\mu\nu}$ (by remapping):
\begin{equation}
r_{0,1}=\sqrt{\int_{S(r_{0,1})} \frac{\Pi}{4\pi}}=\sqrt{\int_{S(r_{0,1})} \frac{\lambda}{4\pi}}\Longrightarrow\int_{\partial \Sigma}(\lambda-\Pi)=0.
\end{equation}
Apart from getting rid of the problematic term on the right-hand side of (\ref{finalResult}), this assumption also makes the factor $r$ in the boundary integral on left-hand side equal to the square root factor in (\ref{Hawking}). If we observe, in addition, that up to second order corrections in $h_{\mu\nu}$:
\begin{equation}
{\rm H}^\mu \rm{H}_\mu \approx k^2 - (K_{AB}\eta^{AB})^2,
\end{equation}
then the left-hand side of (\ref{finalResult}) becomes equal to $16\pi\mathcal{H}_\text{\tiny Haw}$ (the integral of the Ricci scalar is equal to $8\pi$ by the Gauss--Bonnet theorem).

With these observations and assumptions, we may rewrite (\ref{finalResult}) in a cleaned up, final form:
\begin{eqnarray} \nonumber
  16\pi \mathcal{H}_\text{\tiny Haw}
  &\approx & 16\pi \mathcal{H}_\text{\tiny Inv} + \int_{\Sigma} \frac{\Pi}{2r^2} \underbrace{\dip(\y) (-\dtwo)^{-1} \dip (\y)}_{\text{square of angular momentum}} \\ & & \label{finalResultcorrected}
+\frac 12 \int_{\partial \Sigma} \frac{f\Pi}{r} \left( \underline\y \invmd \underline\y  + %\frac{f\Pi}{r}
\underline\x (\B-1) \invmd \underline\x  \right) + \frac{rf}{\Pi} \underline\Xi \invmd \B \underline\Xi \\ \nonumber
& & - \frac 12 \int_{\partial \Sigma}
 \frac{f\Pi}{r} \B \underline{Q} \dtwo\invmd \left[  \underline\x + \frac 14 \dtwo \underline{Q} -\frac 12 f(\B-1)\underline{Q} \right].
\end{eqnarray}
We would now like to interpret the boundary integrals on the right-hand side. The quadratic expressions in $\underline\x$ and $\underline\y$
%in the second line
are constant if we control the value of these ``true degrees of freedom'' at the boundary and can then be neglected: once we calculate the variation of (\ref{finalResultcorrected}), these expressions will turn into terms of the form $\underline\x\delta \underline\x$ and $\underline\y \delta \underline\y$, which vanish if $\delta \underline\x|_{\partial\Sigma} = 0 = \delta \underline\y|_{\partial\Sigma}$. Mathematically, controlling the Dirichlet data on $\partial \Sigma$ for the ``true degrees of freedom'' is necessary whenever we want to describe field evolution within the domain $\Sigma$ as a Hamiltonian system (see \cite{kijowskiGRG,ptcjjk,DeserJPA,KT}). Physically, such control defines an adiabatic insulation of the physical system we want to describe (i.e.~the field contained in the interior of $\Sigma$) from the ``rest of the World''.
%Being constant, the second line in formula \eqref{finalResultcorrected} may thus be neglected.
This belongs to the standard repertoire of the Hamiltonian field theory.

Enforcing the vanishing of the gauge-dependent quantities $\underline\Xi$ and $\underline{Q}$ in the last line seems, at first glance, more problematic. It is, however, justified by the fact that the boundary gauge condition $\underline\Xi = 0 = \underline{Q}$ is the linearized version of the ``rigid sphere'' condition (see \cite{rigid1} and \cite{rigid2}):
A topological sphere $\mathcal{S}$ is a \emph{rigid sphere} if its extrinsic curvature $\rm H^\mu$ is spacelike and the mono-dipole-free parts of the length of $\rm H^\mu$ together with the divergence of its extrinsic torsion vanish:
%\begin{align}
$\displaystyle \underline{||{\rm H}||}=0$,
$\displaystyle \underline{(\lambda t^A)_{||A}} = 0$.
%\end{align}
The spheres $S(r)$ in the Kottler metric satisfy these conditions.
It turns out that the quantities $Q$ and $\Xi$ are linear corrections to the equations above. Therefore, the gauge choice forced upon us by the last line in (\ref{finalResultcorrected}) can be interpreted as a linearized rigid sphere condition. It can also be interpreted as a condition imposed on the vector field $T = \frac{\partial}{\partial t}$, i.e.~on the reference frame.
As noted in the introduction,
 a Hamiltonian functional generating evolution with respect to a generic reference frame does not have the properties which we expect from a true energy, namely positivity, convexity etc. But the quadratic expression $\mathcal{H}_\text{\tiny Haw}$ does fulfill these properties! Thus, the Hawking mass measured on a rigid sphere is positive and convex, at least in the regime of weak fields, when its quadratic approximation prevails. It represents the field energy measured with respect to a reference frame $T$ satisfying the condition:
%\begin{equation}
$\displaystyle g_{\mu\nu}T^\mu {\rm H}^\nu  = 0$.
%\end{equation}
If the boundary of $\Sigma$ does not satisfy the rigidity condition $\underline\Xi = 0 = \underline{Q}$, the corresponding quantity $\mathcal{H}_\text{\tiny Haw}$ is not necessarily positive even in the weak field regime and cannot be identified with the field energy contained in $\Sigma$ because the corresponding field $T$ cannot be treated as a ``time translation'' in any reasonable sense.

\section{Conclusions}
We used our formalism of reduced variables to investigate the behaviour of Hawking mass for weak gravitational perturbations of the Kottler metric.
It turns out that quadratic approximation of the quasi-local mass is related to the gauge-invariant Hamiltonian (the generator of dynamics of linear theory) provided that the boundary of the considered region is composed of spheres satisfying a certain gauge condition. This condition is a linearization of the ``rigid sphere condition'' from the complete, non-linear theory. The rigid sphere condition, in turn, characterizes spheres which provide a physically reasonable ``reference frame'' for defining time translations of a region in spacetime --- which is necessary to talk about energy. This observation supports our hypothesis that the quasi-local energy can be reasonably defined only for 2D surfaces $\partial\Sigma$ which satisfy extra conditions --- those for which such time translations are well defined.
\\
The relation \eqref{finalResultcorrected} contains an extra bulk term: the ``square of angular momentum''. But, due to the vector constraint, this term can be rewritten as a boundary term, see \cite{JJTS}.
%Conditions for the choice of boundary spheres (in linear approximation) correspond to the Rigid sphere conditions in the full theory. This observation suggests that quasi-local mass is perhaps well-defined only for regions with appropriately regular boundary.

\subsection*{Acknowledgments}
This work was supported in part by
Narodowe Centrum Nauki (Poland) under Grant No. 2016/21/B/ST1/00940.
One of the authors (P.W.) was  also supported by a special internal Grant for young researchers,
provided by Center for Theoretical Physics, PAS, Warsaw, Poland.
\

\end{document}